# Superconducting microstrip single-photon detector with system detection efficiency over 90% at 1550 nm

GUANG-ZHAO XU,[1,2,3,4] WEI-JUN ZHANG,[1,2,3,4†] LI-XING YOU,[1,2,3*] JIA-MIN XIONG,[1,2,3] XING-QU SUN,[1,2,3] HAO HUANG,[1,3] XIN OU,[1,3] YI-MING PAN,[1,2,3] CHAO-LIN LV,[1,3] HAO LI,[1,3] ZHEN WANG,[1,3] AND XIAO-MING XIE[1,3]

[1]State Key Laboratory of Functional Materials for Informatics, Shanghai Institute of Microsystem and Information Technology, Chinese Academy of Sciences (CAS), Shanghai 200050, China
[2]Center of Materials Science and Optoelectronics Engineering, University of Chinese Academy of Sciences, Beijing 100049, China
[3]CAS Center for Excellence in Superconducting Electronics, Shanghai 200050, China
[4]The authors contributed equally to this work.
Corresponding author: †zhangweijun@mail.sim.ac.cn; *lxyou@mail.sim.ac.cn

**Generally, a superconducting nanowire single-photon detector (SNSPD) is composed of wires with a typical width of ~100 nm. Recent studies have found that superconducting stripes with a micrometer-scale width can also detect single photons. Compared with the SNSPD covering the same area, the superconducting microstrip single-photon detector (SMSPD) has smaller kinetic inductance, higher working current, and lower requirement in fabrication accuracy, providing potential applications in the development of ultra-large active area detectors. However, the study on SMSPD is still in its infancy, and the realization of its high-performance and practical use remains an opening question. This study demonstrates a NbN SMSPD with a nearly saturated system detection efficiency (SDE) of ~92.2% at a dark count rate of ~200 cps, a polarization sensitivity of ~1.03, and a minimum timing jitter of ~48 ps, at the telecom wavelength of 1550 nm when coupled with a single mode fiber and operated at 0.84 K. Furthermore, the detector's SDE is over 70% when operated at a 2.1-K closed-cycle cryocooler.** © 2020 Optical Society of America

## 1. INTRODUCTION

Superconducting nanowire single-photon detectors (SNSPDs) [1] have been proven as one of the most attractive single-photon detectors, as they provide high system detection efficiency (SDE) [2-5], low dark count rate (DCR) [6], low timing jitter (TJ) [7, 8], high photon count rate (PCR) [9], and broadband sensitivity [10, 11]. To date, SNSPDs have been used in many applications, such as quantum key distribution [12, 13], photonic Boson sampling [14], dark matter detection [15, 16], and satellite laser ranging and detection (LIDAR) [17].

To achieve a saturated internal detection efficiency (IDE), it was believed that the width of the superconducting strip is usually fabricated to ~100 nm, which is the same magnitude as the formed size of a normal domain (referred to "hotspot") after photon absorption [18]. However, a theory proposed by Vodolazov [19] in 2017 predicts that a micron-wide dirty superconducting strip is able to detect a single-photon when it is biased by a current close to the depairing current ($I_{dep}$). In 2018, Korneeva et al. have experimentally shown that the micrometer-wide NbN short bridge can detect a single-photon in a wavelength range of 408-1550 nm [20]. Since then, studies of the superconducting microstrip single-photon detector (SMSPD) have emerged. In 2019, Manova et al. developed NbN SMSPD with an SDE of ~30% at 1330 nm wavelength at 1.7-K operating temperature [21]. In 2020, Chiles et al [22] and Charaev et al [23] reported very large active area of SMSPDs with saturated IDE at 1550 nm at sub-1K operating temperature through very thin amorphous materials (2-3 nm $WSi_x$ or ~3 nm $MoSi_x$). Unfortunately, the SDEs of the reported SMSPDs at the telecom wavelength of 1550 nm are still at a low value (<6%), either due to a low IDE [20, 21] or a low optical absorptance (owing to the use of a very thin film, a low filling factor, or a lack of optical cavity [22, 23]). Furthermore, the IDE of the reported NbN SMSPDs at 1550 nm are still far from saturation [20, 24]. How to realize a high-performance SMSPD that can be operated in a closed-cycle cryocooler is still an opening question. In response, more elaborated works have to be done and more insights to the detection mechanism of SMSPD are required. Numerical simulations based on SMSPDs embedded in an optical cavity are necessary. A proper geometrical configuration to reduce the current crowding effect on sharp turns is needed to bias the microstrip close to its $I_{dep}$, while maintaining a high optical absorptance.

This study reports a He ion pre-irradiated NbN SMSPD that can obtain a nearly saturated IDE at 0.84 K operating temperature, with a 7-nm-thick, 1-μm-wide, double spiral strip configuration and an active area of 50 μm in diameter. Combined with a distributed Bragg reflector (DBR)-based cavity design and a high filling factor ($f$) of 0.8, results demonstrate a simulated absorption efficiency of the microstrip up to ~100% and an experimental SDE of 92.2% at 1550 nm through single mode fiber (SMF) coupling. The detector also exhibits a low polarization extinction ratio (PER) of ~1.03, a low DCR of ~200 cps, and a minimum system TJ of ~48 ps. Operated in a 2.1-K closed-cycle cryocooler, the detector shows a maximum SDE of over 70% at 1550 nm. In addition, the SMSPD is further coupled with a multimode fiber (MMF), where the detector shows a maximum SDE of over 60% and a TJ of ~50 ps.

## 2. DESIGN AND FABRICATION OF SMSPDS

Numerical simulations are performed using a commercial software (COMSOL Multiphysics). Figure 1(a) shows the schematics of the optical stack of SMSPDs, where the microstrips were stacked on top of the DBR substrate [4]. The DBR structure is comprised of 13 periodic $SiO_2/Ta_2O_5$ bilayers in quarter of the central wavelength of 1550 nm, stacked on the top of the Si substrate. Owing to formation of an optical cavity, the absorptance of the microstrips is greatly enhanced. Figure 1(b) shows the simulated optical absorptance as a function of the microstrip thickness, with a fixed strip width of 1 μm and varied $f$ (0.4-0.8). The refractive index of NbN film used here was 4.91+$i$4.67 at 1550 nm, determined by a commercial ellipsometer. A weak influence on absorptance is observed when the strip thickness is greater than 7 nm. A 1-μm-wide microstrip with $f$ = 0.8 demonstrates high absorptance of ~97%. Moreover, for a 7-nm (10-nm)-thick strip, with $f$ ~ 0.92 (0.84), the absorptance could reach to ~100%. Figure 1(c) shows the wavelength dependence of the simulated absorptance, where small dips in absorptance occur in the resonant band (1400-1750 nm, determined at 3 dB cutoff). This behavior is much different with the simulations for the nanowires on the DBR substrate [4], where no dips of absorptance appeared in the resonant band. This may be contributed to some destructive interferences appear in some specific wavelengths because of the narrow spacing between the microstrips (i.e., grating interference effect when the wavelength is larger than the spacing of the grating). In addition, the absorptance of microstrips in the transverse-electric (TE, solid lines) and transverse-magnetic (TM, dashed lines) polarization showed small differences at high f, resulting in a low polarization sensitivity. For example, for $f$ = 0.8 at 1550 nm, the simulated polarization sensitivity (PER = TE/TM) is found to be 97.2%/95.9%~1.01, which was much smaller than the PER (~3-4) of the regular nanowires with $f$ ~0.6 [4].

According to the simulation, the SMSPDs are designed with a fixed 1-μm width and a varied $f$ of 0.4-0.8. To reduce the current crowding effect, the detectors are patterned with a double spiral strip configuration based on previous studies [25, 26]. As a comparison, different geometrical configurations [see Figure 2(a)-(d)] are also designed with the same width on one wafer, including a short micro bridge (called Bridge), a modified double spiral strip (called Spiral-1), a regular double spiral strip (called Spiral-2), and a conventional meandered strip (called Meander). The difference between Spiral-1 and Spiral-2 was the geometry of central parts, due to the different radius of curvature used. The $f$ of the microstrips mentioned is ~0.8 with an active area of 50 μm in diameter or a side length of 50 μm. One limitation of the double spiral strip configuration is that a photon insensitive zone appears in the center, owing to the use of a wider strip to optimize corner curvature. To maximize the coupling efficiency, the detector can be coupled using a lens fiber (small laser beam waist) with an eccentric alignment. Other method on optimizing the device structure to reduce the current crowding effect will be shown in a separate study.

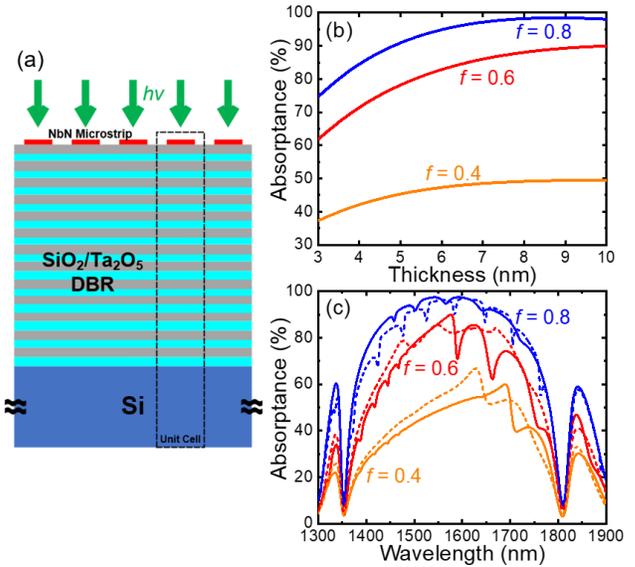

**Fig. 1.** (a) Cross-section schematic diagram of the NbN SMSPD. From top to bottom, optical stacks correspond to a NbN microstrip, a 13-layer $SiO_2/Ta_2O_5$ distributed Bragg reflector, and an Si substrate, respectively. (b) Simulated microstrip thickness dependence of optical absorptance at different $f$ (0.4, 0.6, and 0.8), with a fixed strip width of 1 μm. (c) Simulated wavelength dependence of optical absorptance for microstrips with varied f in a wavelength range of 1300 nm-1900 nm at two different polarizations of light: TE (solid lines) and TM (dashed lines).

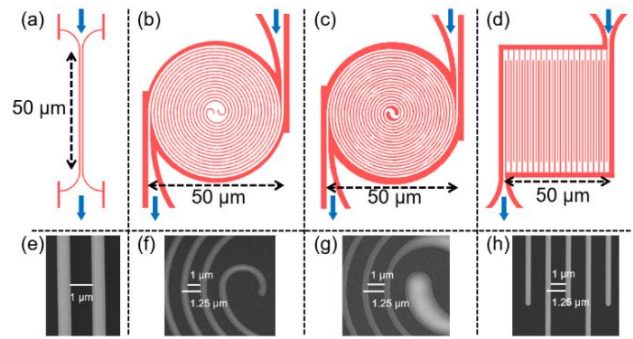

**Fig. 2.** Layouts (top panels, a-d) and magnified SEM images (bottom panels, e-h) of four different SMSPDs. (a) & (e), the short micrometer bridge; (b) & (f), the modified double spiral strip; (c) & (g), the regular double spiral strip; (d) & (h), the conventional meandered strip. The f of the microstrips (b-d, f-h) is 0.8. The blue arrows mark the directions of the current flow.

For fabricating SMSPDs, a 7-nm-thick NbN film is deposited on a 2-inch DBR wafer, using reactive DC magnetron sputtering in a mixture of Ar and $N_2$ gases. To improve the IDE of NbN microstrips, He ion irradiation is conducted to the NbN-covered wafer in a 300-

mm medium-current ion implanter through a He ion energy of 20 keV at room temperature [27]. The ion irradiation fluence was ~5×10$^{16}$ ion/cm$^2$ empirically. Then, the irradiated NbN film was processed to form the designed patterns using electron beam lithography and reactive ion etching (RIE). Figure 2(e)-(h) show the magnified scanning electron microscope (SEM) images of the four different patterns. The coplanar waveguide electrodes were finally fabricated using ultraviolet lithography and RIE.

## 3. MEASUREMENTS AND RESULTS

The SMSPDs are characterized at two different base temperatures: (1) 0.84 K in an adsorption refrigerator and (2) 2.1 K in a compact closed-cycle G-M cryocooler. To prevent the SMSPDs from latching (detector latched at the normal state) [28, 29], a shunted resistor is connected in parallel to the SMSPD chip through wire-bonding. We chose a shunt resistor of ~6.8 Ω (measured at room temperature), which showed optimal performance in IDE and output voltage magnitude. The detector was then biased and read out through a cryogenic coaxial cable, connecting to a DC and RF output port of a bias tee (ZX85-12G-S+, Mini Circuit Inc.) placed at room temperature. Specifically, the bias current was supplied through the DC port of the bias tee, which connected with a series resistor of 20 kΩ and an isolated DC voltage source (SIM928, SRS Inc.). In the RF port, the voltage pulse generated by the SMSPD was amplified using a 50-dB low-noise amplifier (LNA-650, RF Bay Inc.) and then fed into a pulse counter (SR400, SRS Inc.).

Figure 3(a) shows the sweeping current-voltage (I-V) curves for the chip connected with (blue line) or without (red line) a shunt resistor. It can be observed that with a shunt resistor, the nominal switching current ($I_{sw}$) is increased from 66 μA to 80 μA. In the low voltage region (-0.4 to 0.4 mV), the I-V curve demonstrated a slope, which corresponded to a ~5 Ω contact resistance. Because the nominal $I_{sw}$ is influenced by the shunt resistor, we first screened the devices without the shunt resistor. Figure 3(b) shows the $I_{sw}$ comparison of the four different SMSPD configurations on the same wafer with a fabricated width of ~1 μm, measured at 2.1 K. For the $I_{sw}$, at least five samples are tested for each pattern. The average $I_{sw}$s of the Bridge, Spiral-1, and Spiral-2, are 65.4 ± 0.8 μA, 65.2 ± 0.7 μA, and 64.4 ± 1.0 μA, respectively, while that of the Meander is only 43.5 ± 0.9 μA (~0.67 of those of the Spiral-1). Here the symbol "± $x$ μA" of switching currents was referred to a standard deviation, which was estimated from the $I_{sw}$ measurements of different samples. This result confirmed that the sample with a double spiral structure can effectively reduce the current crowding effect, thus guaranteeing a higher $I_{sw}$ (IDE). Therefore, in the following experiment, a modified double spiral strip (Spiral-1) configuration is characterized due to the higher $I_{sw}$.

The optical-electrical performance of the SMSPDs are further characterized based on the reported setup and methods [4]. Specifically, in the SDE measurements, a high-precision optical power meter (81624B, Keysight Inc.) was adopted to calibrate the input power and the attenuation of the attenuators (81570A, Keysight Inc.). A polarization controller was used to adjust the polarization of the input light. We calibrated the input power using the same optical path through switching the input fiber splicing to a fiber jumper connected with the power meter (called monitor port) or to the fiber connected to detector under test (called detector port). Both fiber jumpers were ended with an antireflection-coated facet, which was optimized around 1550 nm to reduce the reflectance (less than 0.3%). A continuous-wave laser (81940A,

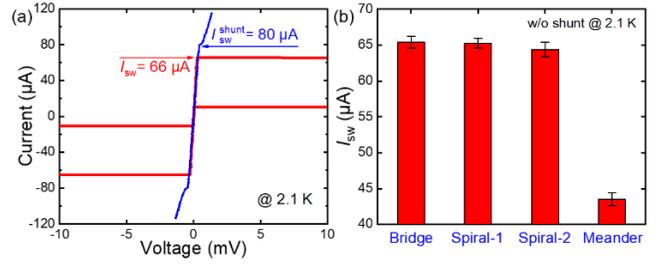

**Fig. 3.** (a) Current and voltage (I-V) trace for NbN SMSPD with (red line) and without (blue line) 6.8 Ω shunt resistor at 2.1 K. The switching currents are 80 μA for shunt ($I_{sw}^{shunt}$) and 66 μA for non-shunt ($I_{sw}$). (b) Switching currents without shunt resistor versus different geometric structures (Bridge, Spiral-1, Spiral-2, and Meander) with error bars measured also at 2.1 K.

Keysight Inc., 1520-1630 nm) was used as the light source. The final input power (−108.92 dBm) corresponded to a photon flux of ~1×10$^5$ photon/s. The power-calibrated fiber jumper was cut and spliced to the detector port. To ensure the power stability, after the measurement, the fiber to the detector port was cut and re-spliced to the monitored port, which showed no obvious changes in the power. The typical spliced loss was less than -0.02 dB, which was included in SDE (a bad splicing would result in a degraded SDE). SDE was determined by the expression of SDE = (CR-DCR)/IPR, where CR is the response count rate, DCR is the dark count rate, and IPR is the input photon rate. The DCR was an average of the CR collected for 10 s when the light was blocked by the shutter.

Then we analyzed the SDE uncertainties of our measurements. Assuming all of the sources of measurement uncertainties were independent, the total measurement uncertainty of SDE was mainly contributed by three factors and could be expressed as: $\sigma_{SDE} = \sqrt{\sigma_{pm}^2 + \sigma_{laser}^2 + \sigma_{att}^2}$. Here, $\sigma_{pm}$ = ±1.81% is the relative uncertainty of the power meter (81624B), calibrated by Physikalisch-Technische Bundesanstalt (PTB); $\sigma_{laser}$ = ±0.09% is the uncertainty of the input laser power (81940A), monitored in a measurement period; $\sigma_{att}$ = ±0.48% is the uncertainty of two cascaded attenuators (81570A). Thus, based on the above parameters, the $\sigma_{SDE}$ was approximately ±1.87%.

Figure 4 shows the comparison of the SDEs versus bias current ($I_b$) for the SMSPDs fabricated with irradiated (called chip "irradiated") and un-irradiated (called chip "un-irradiated") NbN thin films. Both chips have the same film thickness (~7 nm, deposited on the same batch) and the same geometrical configuration (1-μm wide, $f$ = 0.8, and a diameter of 50 μm, Spiral-1 type). The chips were cooled in the 2.1 K G-M cryocooler and were both connected with a shunt resistor and coupled with a lens SMF. The input photon flux was ~1×10$^5$ photon/s at the wavelength of 1550 nm. Owing to the mentioned photon insensitive zone in the center (~10 μm in diameter), the SMF was eccentrically aligned to maximize the coupling efficiency. Notably, the maximum SDEs of the SMSPDs fabricated with irradiated and un-irradiated NbN thin films are ~70% and ~3%, respectively. $I_{sw}$ via irradiation was reduced to ~0.65 of the un-irradiated value, mainly due to the reduction of electron density of states in Femi level ($N_0$) [30].

To explain the significant enhanced IDE of the irradiated SMSPD, the physical parameters of the SMSPDs fabricated with un-irradiated and irradiated NbN thin films (Spiral-1 type) were characterized, as shown in Table 1. It can be found that, the square resistance ($R_{sq}$) was increased and the critical temperature ($T_c$) was suppressed in the irradiated samples, both of which would result in a larger hotspot formation in the microstrip [19]. A larger hot-spot size would help reduce the detection current of the SMSPD. The detection current was referred to a threshold bias current where the absorbed photon drives the superconducting strip to the resistive state [19]. Similar phenomenon was also observed for the irradiated nanowires [27]. Meanwhile, a ratio of $I_{sw}/I_{dep}$ ~0.63 at 2.1 K for the un-irradiated microstrip was deduced, by using the approximate expression of $I_{dep}(T) = 0.74 \frac{w[\Delta(0)]^{3/2}}{eR_{sq}\sqrt{\hbar D}}[1-(\frac{T}{T_c})^2]^{3/2}$ [31]. Here $T$ is the operating temperature, $w$ is the strip width, $\Delta(0) = 1.76k_BT_c$ is the superconducting gap at 0 K, $e$ is the electron charge. The electron diffusion coefficient $D = 1.097\left(\frac{dB_{c2}}{dT}|_{T=T_c}\right)^{-1}$ [32], was estimated from the slope of the curve $B_{c2}(T_c)$ for the SMSPDs with or without irradiation, where $B_{c2}$ is the upper critical magnetic field. This ratio of $I_{sw}/I_{dep}$ for irradiated samples slightly raised to ~0.66 at the same $T$ = 2.1 K. Thus, it is speculated that a combined mechanism may play a role that involves the larger hotspot formation and higher $I_{sw}$ close to the $I_{dep}$ due to ion irradiation effect. Additionally, the results of irradiated samples show that the NbN film currently used in our laboratory is not suitable to achieve high IDE SMSPD at near infrared. Deeper analysis of the changes in the physical properties of the film via irradiation will provide us with guidance for preparing films suitable for SMSPD. Both issues will be explored in another study.

It is worth noting that, SMF-coupling in this experiment had a large tolerance for the misalignment errors, because of the small beam size and the large enough active area. We checked the alignment using an inverted microscope connected with an infrared camera at room temperature. We also confirmed the alignment indirectly by measured the SDE of the detectors. After the cooling system returned to room temperature, we re-checked the alignment and observed no obvious shift of the laser spot.

Empirically, lowering the operating temperature would help improve the IDE of SMSPD. Figure 5(a) shows the temperature dependence of our best irradiated SMSPD coupling with the lens SMF. The SDE (solid scatters) and DCR (open scatters) of the irradiated SMSPD (Spiral-1 type) as a function of $I_b$ are recorded at 2.1 K and 0.84 K, respectively. At 0.84 K, near saturation of SDE appears at the high current region, implying near-unity IDE. A maximum SDE of 92.2% at a DCR of 200 cps are obtained at 1550 nm wavelength. The measured SDE data are fitted at 0.84 K with the sigmoid function (dashed line), showing the saturation trend of the SDE with the current increase. The polarization controller was adjusted to study the polarization sensitivity of detector, as shown in Figure 5(b). The PER (ratio between the maximum and minimum SDEs) of the chip "irradiated" shows a value of less than 1.03, consistent with the simulation (~1.01) at 1550 nm. Low polarization sensitivity is preferred in many applications, e.g., providing a high SDE for the MMF-coupled systems.

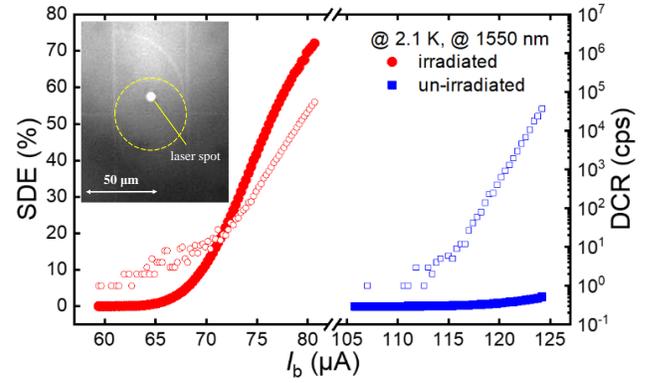

**Fig. 4.** Comparison of SDE (solid scatters) and DCR (open scatters) of the SMSPDs fabricated with irradiated and un-irradiated NbN thin films as a function of bias current ($I_b$) at 2.1 K. The curves were measured with the shunt resistors. Inset: optical coupling image of the tested device captured by an infrared camera after the laser spot (emitted from a lens SMF) eccentrically aligned to the active area of the detector (marked with a dashed circle).

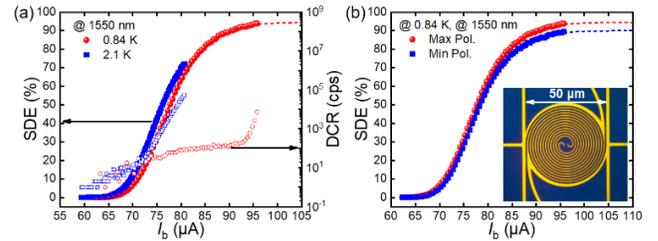

**Fig. 5.** (a) Bias current dependences of SDE and DCR of the SMSPD (chip "irradiated"), measured at two different temperatures, with 1550 nm light illumination. (b) Maximum (solid sphere) and minimum (solid square) SDEs measured at two different polarization of light at 0.84 K. Inset shows a microscope image of the SMSPD with an active area of 50 μm in diameter. Dashed lines are sigmoid function fits in both figures.

Furthermore, we measured the intrinsic DCR of the best irradiated SMSPD with and without shunt. To characterize the intrinsic DCR, the coupled fiber was removed and the chip package block was shielded by aluminum tapes to isolate any optical radiation. It was found that, without shunt, the detector latched and could not produce stable count rate. With shunt, as shown with triangular dots in Figure 6, the intrinsic DCR of SMSPD increased exponentially with current, similar to the behavior of a SNSPD with shunt.

**Table 1. Parameters of the SMSPDs fabricated with un-irradiated and irradiated NbN thin films.** $R_{sq}$(20 K) is the square resistance at 20 K. D is the diffusion coefficient. $I_{dep}$(0 K) and $I_{dep}$(2.1 K) are the calculated depairing current at 0 K and 2.1 K. The $I_{sw}$ at 2.1 K is measured without a shunt resistor

| Samples | $R_{sq}$ (20 K) (Ω/sq) | $T_c$ (K) | D (cm$^2$/s) | $I_{dep}$ (0 K) (μA) | $I_{dep}$ (2.1 K) (μA) | $I_{sw}$ (2.1 K) (μA) | $I_{sw}/I_{dep}$ (2.1 K) |
|---|---|---|---|---|---|---|---|
| un-irradiated | 839 | 7.14 | 0.44 | 185 | 161.5 | 101.1 | 0.63 |
| irradiated | 1036 | 6.40 | 0.50 | 119 | 100.5 | 66.0 | 0.66 |

Note that, the false DCRs was observed in the $I_b>I_{sw}$ region, which was removed from Fig. 6 to avoid misunderstanding. These false DCRs were caused by the RF oscillations due to the use of a shunt resistor [33]. Thus, when the detectors were biased at the same normalized current below $I_{sw}$, we did not observe obvious increase of the intrinsic DCR due to shunt, compared with the DCR of typical SNSPDs with [34] or without [35] shunt. Also similar to the SNSPD, the SMSPD with shunt can be biased at $<0.95 I_{sw}^{shunt}$, where the background DCR (see the open circles in Fig. 6) was at a low level of ~100 cps. Such DCR performance could meet most application needs for low DCR.

Low TJ is a significant advantage of SNSPDs over the other counterpart detectors. It is interesting to determine whether the SMSPD can maintain a low TJ as well as the SNSPD. Previously, TJ in SMSPD showed a strong current dependence and a minimum jitter of ~46 ps was obtained at the current where IDE saturates, measured using a 1064 nm ps laser [21]. Here, we show the system TJ of the chip "irradiated" using the TCSPC module and a 1550-nm fs laser [36]. Figure 7 (a) and (b) show the histogram of the time delay between the laser synchronization signal and output pulse of the SMSPD, recorded at high and low currents at 0.84 K, respectively. TJ was defined as the full width of half maximum (FWHM) of the normalized counts. As shown in Figure 7(a), the count histogram at high $I_b$ of ~95 µA (0.98 $I_{sw}$) was fitted well by the Gaussian distribution, which produced a TJ of 47.5 ps. However, in Figure 6(b), at the lower $I_b$ of ~76 µA (0.79 $I_{sw}$), the TJ increased to 142.4 ps, where count histogram shows non-Gaussian shape with a "shoulder". The "shoulder" can be regarded as the superposition of the main and secondary peaks, as shown by the fit curves [green and orange lines in the Figure 7(b)]. Recent theoretical model has reproduced the non-Gaussian shape by using a modified time-dependent Ginzburg-Landau equation [37]. The mechanism was associated with the position dependent vortex dynamics and the existence of fast and slow absorption sites across the superconducting strip. At the low current, the vortices, and antivortices move slower, leading to increased delay time, thus increasing TJ. Figure 7(c) presents the current dependence of the TJ. Generally, the TJ decreases with the increase of the current. However, at the currents where the IDE changes rapidly (e.g., 72-84 µA, light orange region in the figure), an inflection point of TJ appears in this current region, which may be caused by the effects of the non-Gaussian shape. The arrows in the figure mark two specific currents, at which Figure 7(a) and (b) are reordered.

A minimum system jitter of ~48 ps for our SNMPDs was obtained at $I_b$ = 95 µA. We believe this relatively large system jitter in our experiment was because of the relatively large electrical noise jitter as well as the geometrical jitter. Specifically, because the SMSPD was shunted with a small resistor, the output pulse amplitude was significantly reduced from 2 V to ~190 mV, resulting in a lower slope of the rising edge of the response pulse. This produced a relatively large electrical noise jitter with a magnitude of ~17-40 ps [36, 38]. Besides, the 50 µm diameter active area would produce a geometrical jitter with a magnitude of ~11-25 ps [39][40]. In future, it would be interesting that, exploring the physical limit of the time jitter of SMSPD using cryogenic amplifiers and shorter strip.

Figure 8 shows more details of the chip "irradiated". Figure 8(a) shows the photon-response pulse of the SMSPD, with a fitted decay time (1/e criterion) of ~36 ns for the falling edge of the pulse. Although shunted with a resistor, a high pulse magnitude of ~190 mV was observed, guaranteeing a good signal-to-noise ratio of the output pulse. Figure 8(b) shows the CR dependence of the SDE measured at $I_b$ = 93 µA at 0.84 K. A CR of ~5.7 MHz at the 3-dB cutoff point of the SDE was obtained, while a maximum CR (MCR) of ~15 MHz was determined at the SDE of 10%. The measured MCR was generally less than the MCR deduced from the decay time [1/ (36 ns) ~27 Mcps], possibly owing to the limitation of our current-used AC-coupled readout circuit [40]. However, the advantage of SMSPD is that when the active area is large, there is no notable overshoot effect in the falling edge of the pulse caused by large kinetic inductance [41]. Figure 8(c) shows the wavelength dependence of the SDE at TE polarization at $I_b$ = 93 µA at 0.84 K. At the wavelength range of 1520 nm to 1630 nm, the SDE shows a value greater than 88%. Because the difference between the peak and dip values of the simulated absorptance in this wavelength range is ~2.6%, which is close to the measurement error of the SDE, it is difficult to observe a clear dip (around 1570 nm) in the SDE. When the wavelength is longer than 1590 nm, the SDE demonstrates a slight decrease with the increase of the wavelength because of the non-saturation of the SDEs at the longer wavelength.

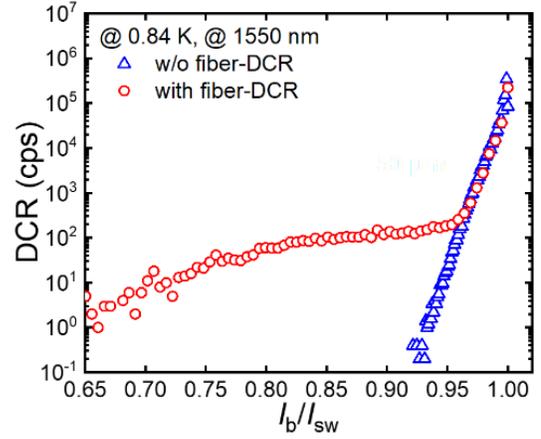

**Fig. 6.** DCR of the best SMSPD with and without fiber (i.e., the intrinsic DCR) as a function of the normalized bias current ($I_b/I_{sw}$), recorded at ~0.84K.

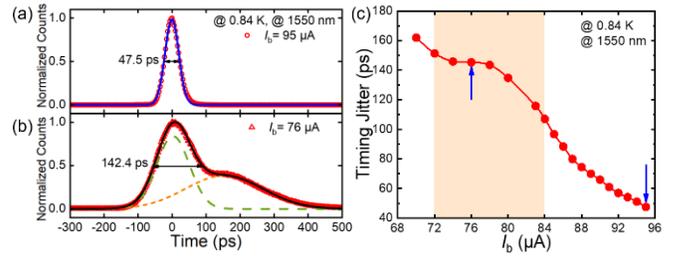

**Fig. 7.** Histogram of time-correlated photon counts measured at 1550 nm: (a) $I_b$ = 95 µA (red circle). The blue line is the Gaussian distribution fit, with the FWHM of 47.5 ps. (b) $I_b$ = 76 µA (red triangle). The black line is superposition of two peaks with the FWHM of 142.4 ps. The green-dashed and orange-dotted lines are the Gaussian distribution fits for the main peak and secondary peak, respectively. (c) The bias current dependence of the TJ in a range of 70-95 µA.

Figure 8(d) demonstrates the performance of our device coupled with a lens MMF with a core diameter of 50 μm and a beam waist of ~28 μm. The MMF-coupled SDE vs. $I_b$ was recorded at two different photon fluxes ($1\times10^6$ and $1\times10^5$ photon/s, respectively). It was found that, the SDE recorded at the low photon flux ($1\times10^5$ photon/s) was fluctuating at the high bias current region (>85 uA), due to the fluctuation of the large DCR. A sigmoid fit was plotted against these experiment data, showing the trend of the SDE($I_b$) curve. The maximum SDEs under these two photon fluxes were ~61% and 63%, respectively, determined at $I_b$~95 μA. The slight increment of ~2% in SDE confirmed there was a weak blocking effect at high count rate. However, the maximum SDE was still lower than expectation. We speculated the relatively low SDE of the MMF coupling in this experiment was mainly attributed to a relatively large misalignment of the laser spot due to the lack of clear alignment marks in the field of view. In future, we would fabricate auxiliary alignment marks on the SMSPDs, similar to what we have done in SNSPDs [4], which would further improve the alignment accuracy. However, according to our knowledge, this SDE is still the highest value reported for the MMF-coupled detectors at 1550 nm. Meanwhile, owing to the broadband background radiation transmitted by the MMF coupling, a significantly raised DCR was observed, which can be suppressed using cold narrowband filters, e.g., a MMF-coupled filter bench [42]. Through Gaussian fitting, TJ of ~50 ps at $I_b$ = 95 μA is obtained, which is slightly larger than that of the SMF coupling due to the fiber-associated dispersion in optical signal transmission in MMF [43].

Finally, the SMSPD performances are compared with the state-of-the-art of the SNSPDs at 1550 nm wavelength listed in Table 2, showing the potential of the SMSPD. The SNSPD with an active area of 50 μm and operated at 1550 nm usually demonstrates a very large kinetic inductance (i.e., a long decay time over 1 μs without a series resistor), large TJ, and a very low yield (based on our own experience for NbN detectors). In contrast, the SMSPD with the same size exhibits an improved decay time, TJ, and yield [~68% (14/21) in one wafer, with a criterion of $I_{sw} \geq 60$ μA without shunt], making it attractive for applications requiring a large active area, high timing performance, and efficient detection.

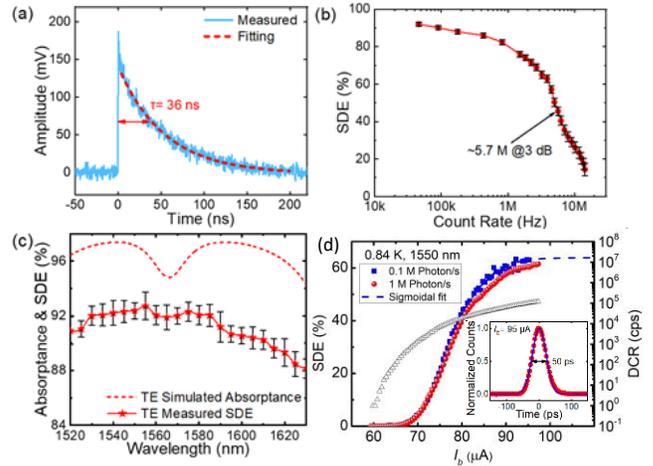

**Fig. 8.** (a) The oscilloscope single pulse waveform graph of response versus time. The exponential fitting of the falling edge is given as 36 ns. (b) The dependence of SDE and count rate of SMSPD at 0.84 K. The count rate is ~5.7 MHz at 3 dB point. (c) Wavelength dependencies of the absorptance and SDE at TE polarization and 0.84 K for simulated absorptance (red dashed line) and the measured values with error bars (red stars). (d) The SDE and DCR versus $I_b$ with an MMF coupling at 0.84 K, recorded at two different photon fluxes: 0.1 M photon/s (blue square), 1 M photon/s (red sphere). Dashed line is the sigmoid fit for the data recorded at 0.1 M photon/s. Inset shows the MMF coupled TJ is 50 ps at $I_b$ = 95 μA.

**Table 2. Comparison of the key merits of the SNSPDs and SMSPDs operated at 1550 nm wavelength**

| Detectors | Material | Area (μm²) | Width (nm) | SMF Coupling | | | | MMF Coupling | | | Decay time (ns) |
|---|---|---|---|---|---|---|---|---|---|---|---|
| | | | | SDE (%) | DCR (cps) | TJ (ps) | PER | SDE (%) | DCR (cps) | TJ (ps) | |
| SNSPD | MoSi$_x$[3] | Φ50 | 80 | 98.0 | ~10² | ~550 | 1.23 | N/A | N/A | N/A | ~400 |
| | WSi[5] | Φ15 | 120 | 93.2 | ~10³ | 150 | 1.16 | N/A | N/A | N/A | ~75 |
| | NbN[4] | Φ15 | 75 | 92.1 | ~10 | 40 | 3.5 | N/A | N/A | N/A | ~27 |
| | NbTiN$_x$[43] | Φ50 | 70 | 75 | ~10² | 18.7*@ 1.06 μm | 3.75 | 50 | ~10⁵ | N/A | N/A |
| SMSPD | MoSi$_x$[22] | 400×400 | 1000 | <6 | ~10² | N/A | N/A | N/A | N/A | N/A | ~75 |
| | WSi[23] | 362×362 | 2000 | N/A | ~10³ | N/A | N/A | N/A | N/A | N/A | ~45 |
| | NbN[21] | Φ20 | 1000 | 35@ 1.3 μm | ~10⁴ | 45 | N/A | N/A | N/A | N/A | 2.5** |
| | NbN (this paper) | Φ50 | 1000 | 92.2 | ~10² | 47.5 | 1.03 | 63 | ~10⁵ | 50 | 36** |

*Use of a low temperature amplifier.
**Not identical to the rest time, due to the influence of the shunt resistor.

## 4. DISCUSSIONS

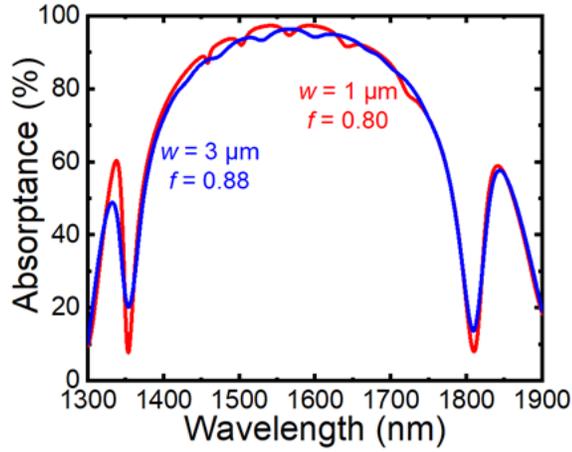

**Fig. 9.** Simulated wavelength dependence of optical absorptance for microstrips with two different widths ($w = 1$ and $3$ μm) and two different filling factors ($f = 0.80$ and $0.88$) in a wavelength range of 1300-1900 nm, respectively.

Here we provide more insights and discussions to our results. Firstly, for optical cavity design, recent simulation and experiment results (e.g., [2, 4] and this paper) have shown that, near unity absorptance and SDE can be obtained without the additional layers stacked on the top of the NbN strip, because of the formation of a strong half-wave cavity. Adding the additional layer on the top of the NbN strip would result in a narrower resonated bandwidth but with no obvious enhancement in absorptance. Therefore, to simplify the fabrication process, we did not fabricate the additional dielectric layers. Secondly, for the strip geometry, we have simulated strips with varied widths (e.g., 1-3 μm) and varied filling factors (e.g., 0.3-0.97). It was found that, each strip width was possible to achieve a maximum absorptance close to 100%. For instance, we show a comparison of the simulated absorptance for micro strip with widths of 1 and 3 μm at specific filling factors in Figure 9. A maximum absorptance over 95% was both obtain at these two strip geometrical structures. Therefore, the choice of strip width mainly depends on the processing accuracy and the actual requirements. In this study, the use of 1 μm wide strips with a varied $f$ of 0.4-0.8 was empirical. Thirdly, the double spiral strip configuration was useful to improve the maximum bias current. However, as pointed out earlier, it has a drawback that it is insensitive to detection in the middle, thus requiring more sensitive alignment and results in wasting of the SMSPD active area. Thus, we thought the spiral strip configuration would be useful for the case in which the SMSPDs have a very large active area and are coupled to a beam with a large beam size. Thus, the middle insensitive area would sacrifice a very small part of SDE. For example, considering a very-large active detector with a 10-μm diameter insensitive area coupled to a 150-μm diameter Gaussian beam, the estimated coupling loss was ~0.9%. Besides, recent advances in reducing the effect of current-crowding have been made, such as thickening the turns of the meander strip [44].

In our experiments, the SMSPD was shunted with a resistor, which would cause the RF oscillations at the $I_b$ greater than $I_{sw}$ (up to ~1.4$I_{sw}$, empirically), where the detector was suffered from the RLC oscillations. This phenomenon was also observed in the SNSPD shunted with a resistor [33]. In this current region, the pulse waveform of the RF oscillations observed by the oscilloscope was stable and repeated at a specific frequency, which was easy to distinguish from the normal photon response of the detector. Furthermore, in the DCR measurement, the DCR logarithmically increases with the bias current. However, after entering the RF oscillation region, the curve of the DCR vs. $I_b$ would show a different slope, which provides evidence for us to distinguish the normal operation region from the RF oscillation region. It is worth noting that, generally, our detectors are biased below the switching current, and the RF oscillations would not affect the detector's performance. Moreover, in the low bias current region (below $I_{sw}$), we also did not observer obvious RF oscillations induced by the shunt (which implied by a theoretical prediction in [37]), through monitoring the periods of the pulse of the dark count. We speculated this type of RF oscillations either possibly occurred at specific kinetic inductance, shunt resistor, and strip width, or too weak to observe.

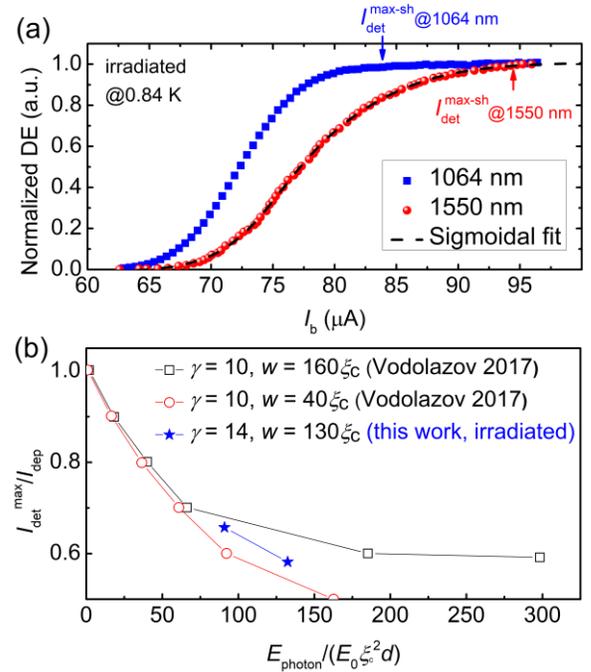

**Fig. 10** (a) Bias current dependence of normalized detection efficiency (NDE) for the irradiated SMSPD, recorded under the 1064 nm and 1550 nm photon's illumination, respectively, while operated at 0.84K with a shunt resistor. Arrows indicate the locations of the $I_{det}^{max\text{-}sh}$ determined for the two different wavelengths. Dashed line is a sigmoidal fit for the data measured at 1550 nm. (b) Dependence of the maximal detection current on the photon's energy at different $\gamma$ and width ($w$). The open dots were the calculated data obtained from Vodolazov's paper (Inset of Fig. 11 [19]), with $\gamma = 10$, $\xi_c = 6.4$ nm, and $T_c = 10$ K for NbN. The solid star symbols are our experimental results, with estimated $\gamma = 14$, $\xi_c = 7.7$ nm, and $T_c = 6.4$ K for the irradiated NbN device.

In terms of detection mechanism, we adopted a simplified diffusion hotspot model [45] to further explain the SDE enhancement caused by ion irradiation. This model describes the photon response of a superconducting strip through an analytical expression of $E_{min} = \frac{hc}{\lambda_{max}} \geq \frac{N_0 \Delta(0)^2 wd}{\varsigma} \sqrt{\pi D \tau_{th}} \left(1 - \frac{I_b}{I_{dep}}\right)$, where $E_{min}$ is a minimum energy (or corresponding to a cut-off wavelength $\lambda_{max}$) detectable by the superconducting strip, $c$ is the speed of light, $\tau_{th} \approx 1.6$ ps is the time scale of the quasiparticle multiplication process and $\varsigma \approx 0.25$ is multiplication efficiency of quasiparticles

[46]. According to the data of our electrical transport measurements, the ion irradiation was found to affect the related physical parameters of the strip (e.g. $N_0$, $\Delta$, $D$, $I_b/I_{dep}$), then the $E_{min}$ due to irradiation was estimated using the change ratio of these parameters. Via calculation, $E_{min}$ was reduced to nearly 65% of its un-irradiated value. Correspondingly, the $\lambda_{max}$ were extended to longer wavelength which implied an enhanced spectral sensitivity.

Recent theoretical works have extended the analytical hotspot model to more complicated models [19][47], which require numerical simulations. For example, the calculation by Vodolazov [19] based on a hotspot tow temperature model (assuming a short thermalization time at the initial stage of the hotspot formation) predicted the single-photon detection ability of the strip with a micro scale width, when the maximal detection current ($I_{det}^{max}$) exceeded a specific ratio of $I_{dep}$. Here, we compared the detection current of our devices with the theoretical results shown in Vodolazov's paper [19]. Firstly, we determined the $I_{det}^{max}$ of our device through measuring the normalized detection efficiency (NDE) as a function of $I_b$, illuminated at a specific photon energy (wavelength). As shown in Fig. 10(a), the data of the irradiated device were recorded at 0.84 K and illuminated at two different wavelengths (1064 nm and 1550 nm, respectively). From the NDE$(I_b)$ curves, the maximal detection current ($I_{det}^{max-sh}$) with shunt was defined as the current at which the NDE became greater than a threshold value of 0.99 (i.e., the NDE became saturated). Assuming the $I_{dep}^{sh}$ have the same increase ratio (~1.21, empirically) as the $I_{sw}^{sh}$ due to shunt, then the $I_{det}^{max-sh}$ was normalized to the $I_{dep}^{sh}$, i.e., $I_{det}^{max-sh}/I_{dep}^{sh}$. We further assumed the $I_{det}^{max-sh}/I_{dep}^{sh} = I_{det}^{max}/I_{dep}$.

Then, according to Vodolazov's paper [19] and our measured physical parameters (shown in Table 1) for the irradiated NbN device, we calculated the relevant physical parameters of our device in terms of Vodolazov's paper [19]: $\xi_c = \sqrt{\hbar D/k_B T_c} \approx 7.7$ nm, the characteristic energy of $E_0\xi_c^2 d \approx 8.8$ meV, where $E_0 = 4N_0(k_B T_c)^2$, and the thickness $d = 7$ nm. We also estimated the coefficient $\gamma \approx 14$, $w = 1000$ nm $\approx 130\xi_c$, and the excitation photon engorge of $E_{photon} = hc/\lambda \approx 0.80$ (1.17) eV at 1550 (1064) nm, respectively.

Based on the mentioned above parameters, especially for the value of $\gamma \approx 14$ in our NbN device, we compared the experiment data ($I_{det}^{max}/I_{dep}$ vs. $E_{photon}/E_0\xi_c^2 d$) with the calculated data took from Vodolazov's paper [19], as shown in Figure 10(b). Two sets of the calculated data which corresponded to two different strip widths ($w = 160\xi_c$ and $40\xi_c$) with the same of $\gamma \approx 10$ were plotted against our data. Because the $\gamma$ and $\xi_c$ values of these data are close, the strip width would play a key role on the ratio of $I_{det}^{max}/I_{dep}$, when the strip illuminated with the same photon energy. Thus, the two set of the calculated data may serve as the upper and lower boundaries for our results. However, more experiment data are needed to draw a full picture of the curve, especially for the low energy region (corresponded to the longer wavelength > 1550 nm) and high energy region (corresponded to the shorter wavelength <1064 nm). These works would be done in the later experiments. It is also interesting that, in the low energy region (e.g., $E_{photon}/E_0\xi_c^2 d < 60$), whether the experiment data for different strip widths would overlap each other; while in the high energy region (e.g., $E_{photon}/E_0\xi_c^2 d > 200$), whether the ratio of $I_{det}^{max}/I_{dep}$ would tend to saturation. Both of these studies would provide more information to the understanding of detection mechanism.

## 5. CONCLUSIONS

In conclusion, this paper simulated, fabricated, and characterized a NbN microstrip on a DBR substrate with various filling factors (0.4-0.8) and various strip configurations (bridge, double spiral, and meander). Simulation shows that a high filling factor is necessary to achieve high SDE in the SMSPD. A double spiral strip configuration is helpful in reducing the current crowding effect. Owing to the use of the NbN film pre-irradiated by He ions, the IDE of the NbN SMPSD is significantly improved, providing more physical insights to the detection mechanism of the SMPSD. Based on the abovementioned methods, this study successfully demonstrated the NbN SMSPD with a strip wide of 1 μm, a filling factor of ~0.8, and an active area of 50 μm in diameter, showing a maximum SDE of 92.2% at 1550 nm, a DCR of 200 cps, a minimum TJ of 48 ps, and a PER of 1.03 at 0.84 K. Operated in a 2.1 K closed-cycle cryocooler, the detector shows a maximum SDE of over 70% at 1550 nm. In addition, the SMSPD was further coupled with a multimode fiber, where the detector shows a maximum SDE of over 60% and a TJ of ~50 ps. Results of this study shed light on the development SMSPDs for efficient single-photon detection, which would show the potential applications prospects in quantum optics and photon-starved LIDAR.

**Funding.** This work is supported by the National Natural Science Foundation of China (NSFC, Grants No. 61971409), the National Key R&D Program of China (Grants No. 2017YFA0304000), and the Science and Technology Commission of Shanghai Municipality (Grants No. 18511110202 and No. 2019SHZDZX01). W. J. Zhang is supported by the Youth Innovation Promotion Association, CAS (2019238).

**Acknowledgments.** The authors thank Xiaoyu Liu for technical assistance in EBL, and Peng Hu for technical assistance in detector's optical coupling. The authors also thank Huiqin Yu for technical support in the use of 0.8-K cryostat.

**Disclosures**. The authors declare that they have no conflicts of interest.